# Reducing Toxic Thallium in Tl-2223 Cuprate

R. Shipra,[1,2] A.S. Sefat[2]

[1]Department of Physics and Astronomy, Vanderbilt University, Nashville, Tennessee 37235, USA

[2]Materials Science and Technology Division, Oak Ridge National Laboratory, Oak Ridge, Tennessee 37831, USA

**Abstract**

The present study gives an account of the effect of $Li_2O$ on the ease of phase formation and superconducting properties of $Tl_2Ba_2Ca_2Cu_3O_{10+\delta}$ (Tl-2223) material. Li substitution only slightly decreases the superconducting transition temperature, while the optimal concentration of 20% doping improves the critical current density ($J_c$) by ~ two fold. We found substantial effects on the synthesis temperature, microstructure and normal state transport properties of Tl-2223 with $Li_2O$ addition. Short-time thermal annealing under flowing Ar+4%$H_2$ (1 hr.) further improves the superconducting volume fractions, as well as $J_c$.

**Introduction**

High volatility and toxicity of $Tl_2O_3$ makes the material preparation of thallium-based cuprate superconductors complex, because slight variation in the synthesis condition can give variable oxide phases. Tl-2223 cuprate has a high superconducting transition temperature of $T_c \approx 125$ K comparable to $TlBa_2Ca_2Cu_3O_{9-\delta}$ (Tl-1223), but has a better phase stability during synthesis at ambient pressure.[1,2,3] In Tl-2223, Tl exists in both +3 and +1 oxidation states and so, their relative amounts can vary with oxidation and reduction annealing conditions. The coordination number of Tl is 6 with oxygen, while Ca and Cu are coordinated to 8, and 4 or 5 oxygens, respectively. Some reports emphasize the role of excess Ca and Cu in stabilizing this phase.[4,5] In this work, our basic hypothesis was that lithium can substitute in place of toxic thallium to the extent that 2223 structure remains stable and most superconducting properties remain intact.

Although we have nominally accounted for Li on Tl atomic site, the small $Li^+$ can potentially substitute at any combination of the cationic sites[6], or even interstitial sites, influencing crystal structure and hole concentration. There is no literature report of such chemical substitution in Tl-2223; however, there are a few reports on Tl-2212 and the Tl-site substitution[7,8] by $Rb^+$ and $K^+$ with similar sizes to $Tl^+$.[7,8] In fact, Li substitution may be favored at $Cu^{+2}$ and $Ca^{+2}$ sites; for example, it is found that Li at Ca site drastically decreases $T_c$, but when substituted for Cu, the $T_c$ decrease is slow for the same amounts of substitution in Bi-2212.[9] Other reports have suggested an optimum amount of Li concentration (~ 20% per formula unit) to improve the $T_c$ and $J_c$ of Bi-2212, Hg-1223, (Tl,Pb)-1223.[10,11,12,13] Reports on less (structurally, chemically) complex cuprates like $La_{2-x}Sr_xCuO_4$ (LSCO) and $YBa_2Cu_3O_{7-\delta}$ (YBCO) provided additional insight on the effect of Li substitution at the Cu-site.[14,15,16] For example, the non-magnetic $Li^+$ impurity was found to suppress $T_c$ more rapidly as compared to magnetic impurities like $Ni^{+2}$.[17,14] In melt textured YBCO, $Li^+$ was found to act as effective pinning centers at optimal concentrations.[18]

Though these reports provide some understanding of the effect of chemical substitutions, no clear trend in the nature of doping can be concluded. In this work, we provide a simple route to obtain polycrystalline Tl-2223 superconductors with high reproducibility and present a way to improve its superconducting properties with the addition of $Li_2O$ and thermal annealing. In the past, long time thermal annealing in an evacuated sealed quartz tube has shown to play an important role in increasing $T_c \approx 128$ K.[19,5,20] However at such high annealing temperatures, no discussion on the effect of Tl loss and inter-site cation substitution on $T_c$ was discussed.[20] In this study, we did short time (1 hour) thermal annealing at a low temperature (250 °C) under flowing Ar+4%$H_2$, as longer duration resulted in the formation of Tl-free impurity phases. The annealing condition was carefully chosen and based on the results obtained by Maignan *et al* for Tl-2201.[1,21] The focus on this study is to add knowledge regarding the effect of Li on the crystal structure, microstructure, normal and superconducting properties of Tl-2223.

**Experimental**

For synthesizing polycrystalline samples equivalent to $Li_xTl_{2(1-x)}Ba_2Ca_2Cu_3O_{10}$ ($Li_x$Tl-2223), first a precursor was prepared having a composition equivalent to $Ba_2Ca_2Cu_3O_7$ and then appropriate molar amounts of $Tl_2O_3$ and $Li_2O$ were added to complete the formation of the desired compound, with x = 0, 0.1, 0.2, 0.3, or 0.4. First of all, $BaCO_3$, $CaCO_3$ and CuO were ground in the cationic ratio of 2:2:3, and calcined subsequently three times at 880 °C, 890 °C, and 900 °C, respectively in air, with homogenizing grinding steps in between. Each calcination was done for 12 hrs. The final mixture was pelletized at a pressure of 6000 tonnes/m$^2$ and sintered at 910 °C in air for 48 hrs. The X-ray diffraction (XRD) pattern showed the mixture (precursor) consisting of $BaCuO_2$, $Ca_2CuO_3$ and CuO, to which $Tl_2O_3$ and $Li_2O$ were then added. The best optimized ratio of $Tl_2O_3$:precursor for preparing the single phase Tl-2223 was found to be 1.1:1.

For synthesizing $Li_{0.1}$Tl-2223 and $Li_{0.2}$Tl-2223, the ratio of $Li_2O$:$Tl_2O_3$:precursor were mixed in the molar ratio of 0.05:1.05:1 and 0.1:1:1, respectively. The mixed powders were ground well, and pressed by applying a pressure of 1500 tonnes/m$^2$ into ¼ inch single pellets each weighing between 0.5-0.75 grams. Each pellet was wrapped in a gold foil, sealed in a quartz tube individually under 0.9 atm. of oxygen, and then heated at 910 °C for 3 hrs. For synthesizing $Li_{0.3}$Tl-2223 and $Li_{0.4}$Tl-2223, $Li_2O$:$Tl_2O_3$:precursor were mixed in the ratio of 0.15:0.95:1 and 0.2:0.9:1. The pellets were sintered at 900 °C, also in sealed quartz tube with 1 atm. of oxygen. We must note that neither phase formed at 910 °C with XRD patterns showing numerous impurities, hence large sensitivity of phase formation to small temperature differences. The 'as-prepared' pellets were broken to retrieve pieces for 'annealing' studies (Ar+4%$H_2$ for 1 hr). The data of 10, 20, and 30% nominal Li substitutions for Tl are shown below.

XRD patterns were obtained using X'pert PRO MPD powder diffractometer in the 5-90° 2θ range. The refined cell parameters were determined from Lebail refinements/matching of the experimental data (reference code: 01-081-0044) in the X'pert HighScore Plus software. Zero shift, unit-cell dimensions and profile values are the only parameters that were refined. The $T_c$ and $J_c$ were determined from magnetic measurements performed on SQUID magnetometer. The zero-field-cooled (ZFC) data were collected while increasing the temperature from 5 K to 150 K while the field-cooled (FC) data were obtained while cooling from 150 K in field. For a few samples we determined the resistive transition by transport measurements with Physical Property

Measurement System (PPMS) by Quantum Design (USA). For estimating $J_c$, field-dependent isothermal magnetization hysteresis curves were obtained at 5 K.

**Results and discussion**

Figure 1 shows the XRD pattern of the as prepared $Li_xTl$-2223 samples. Tl-2223 sample contained $BaCO_3$ (shown as #) as the only impurity while addition of $Li_2O$ promoted Tl-2212 and $BaCuO_2$ (both marked as *) impurity phases. Well-resolved XRD peaks indicate good crystallinity and can be indexed accordingly. The lattice parameters obtained for Li free Tl-2223 were $a$ = 3.8500(7) Å and $c$ = 35.6392(8) Å. With increasing Li content, $a$ lattice parameter showed a slight increase while $c$ and $c/a$ ratio decreased fast (almost linearly). The lattice parameters obtained for x = 0.4 ($Li_{0.4}Tl$-2223) were $a$ = 3.8507(1) and $c$ = 35.603(1). Similar trend of $c$ lattice parameter decreasing with Li addition was reported for Bi-2212.[9] For LSCO, $a$-lattice parameter along with total volume decrease, while the $c/a$ increases when Li presumably substitutes Cu.[22] After annealing Tl-2223 in Ar+4%$H_2$, $a$ = 3.84972(8) Å while a slight increase in $c$ (35.656(1) Å) was observed. Similar increase in the $c$-axes were observed in all other Li added samples. It is to be noted that no extra impurity peak appeared in the XRD patterns after annealing.

Figure 2 shows the scanning electron micrographs (SEM) of the as prepared $Li_xTl$-2223 samples. The average grain size for lithium-free Tl-2223 (x= 0) is less than 20 μm, which seems to increase in Li incorporated samples of x=0.1, 0.2, 0.4. This trend is similar to that observed in Li doped (Tl,Pb)-1223 and Bi-2212 superconductors.[12,11] For $Li_{0.1}Tl$-2223, clear evidence of spiral growth due to the presence of screw dislocations, similar to that observed in YBCO, leading to slab like grains.[23] Such morphologies may lead to specific grain orientations causing a degree of alignment. Li incorporated pellets were much shinier in appearance compared to Li free Tl-2223.

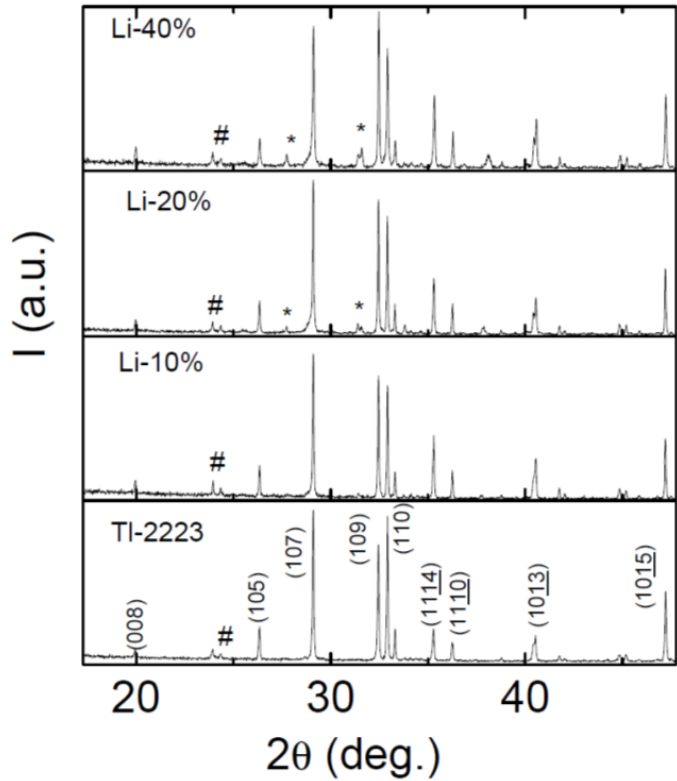

**Figure 1**: X-ray diffraction patterns of the as-prepared samples with nominal formula of $Li_xTl_{2(1-x)}Ba_2Ca_2Cu_3O_{10}$ ($Li_xTl$-2223), where x = 0, 0.1, 0.2 and 0.4. Bragg peaks contributing to phases of $BaCO_3$ and Tl-2212 are denoted as # and *, respectively. All peaks are normalized to the (110) peak position.

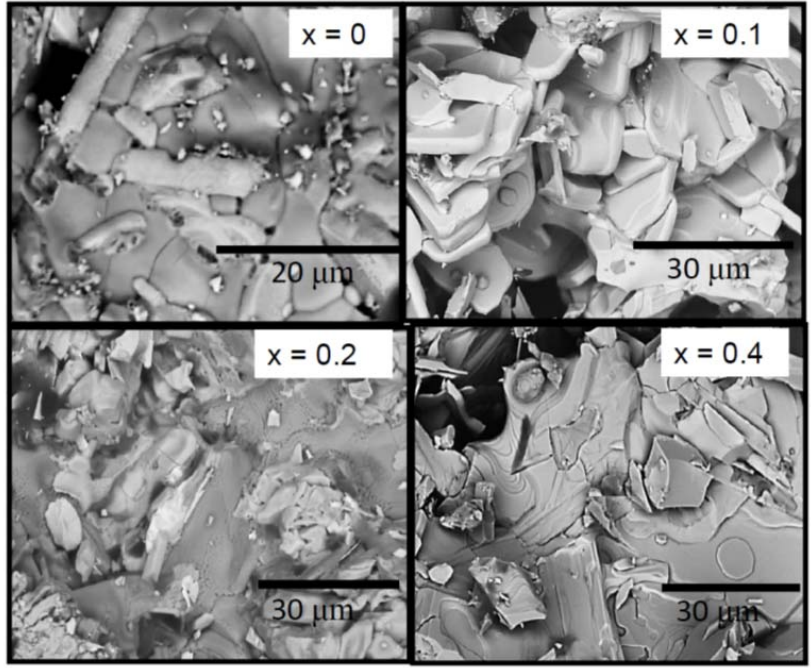

**Figure 2**: Scanning electron micrographs of pure and $Li_xTl$-2223. Li addition seems to improve microstructure by increasing the grain size.

Figure 3 is the temperature-dependent magnetic susceptibility, showing ZFC (open symbols)-FC (closed symbols) curves of $Li_xTl$-2223 samples. The $T_c$ values for all samples are approximately the same. For as-prepared Tl-2223 (Fig. 3a), ZFC curve shows an upturn (reproducible feature) below 75 K. With Li doping, the upturn gets suppressed and the screening as well as Meissner fractions improve, showing best results for $Li_{0.2}Tl$-2223. Annealed Tl-2223 sample improves with sharper diamagnetic transition (Fig. 3b), while there is not much change or trend within $Li_xTl$-2223. Long-time annealing (~5 hrs) at 250 °C decreased the $T_c$ as well as the superconducting volume fraction of Tl-2223, and therefore the data are not presented here. It is also evident that $T_c$ only slightly decreases with the addition of lithium, similar to that reported for (Tl,Pb)-1223.[12] Li free Tl-2223 shows a diamagnetic transition at ~ 124 K, while the value decreases to ~ 122 K in $Li_{0.1}Tl$-2223, and ~ 119 K in $Li_{0.4}Tl$-2223. Such a decrease in $T_c$ and change in the structure and morphology may indicate that Li gets into the lattice; however exact location of Li cannot be inferred from these any of such observations. Not much change in the $T_c$ from magnetization was observed after annealing in Ar+4%H$_2$.

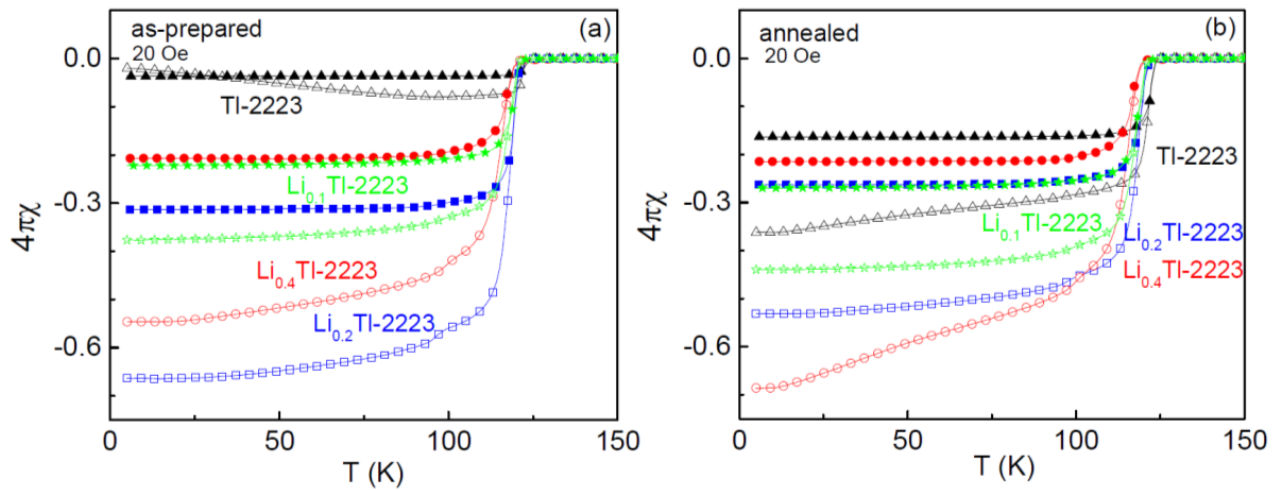

**Figure 3**: Temperature-dependent magnetic susceptibility results measured under 20 Oe assuming perfect diamagnetism. The ZFC curves are closed symbols, while FC curves are open symbols.

For estimating the $J_c$ in these materials, magnetic hysteresis curves were obtained at 5 K. Shown in Figure 4 is the $J_c$ of Li$_x$Tl-2223 extracted from the magnetic hysteresis curves for the lithium added samples using the Bean's critical state formula for rectangular sample: $J_c = 20\Delta M/[b(1-b/(3a))]$, where $a$ and $b$ are dimensions of the cross-section of $a > b$.[24] Here $\Delta M$ is the hysteresis width given as, $\Delta M = M^- - M^+$, where $M^-$ is the magnetization in field decreasing part and $M^+$ is the magnetization in the field increasing part of the hysteresis loop. Lithium added samples show higher $J_c$ as compared to pure Tl-2223, with Tl$_{0.2}$Tl-2223 showing the highest $J_c$ in both the as-prepared and annealed samples. These results match well with previous reports where $J_c$ increases with increasing Li content and reaches a maximum for the nominal optimal value of 20%.[12] Though this work does not explain the position of Li in the lattice, there is a possibility that if substituted at Cu in optimal amounts, it can act as strong vortex pinning site without affecting the $T_c$.[18] Here, increase in the grain size and granularity may adversely affect the increase in $J_c$.

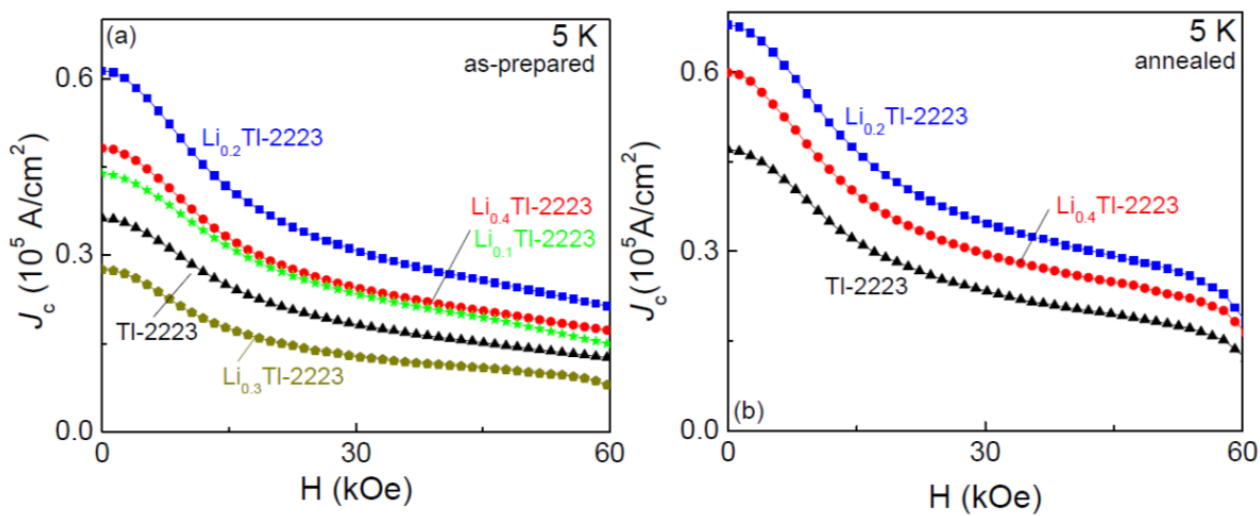

**Figure 4**: Field-dependence of current densities of the (a) as-prepared and (b) annealed Li$_x$Tl-2223 at $T = 5$ K, calculated using the Bean's model.

Figure 5 shows the resistivity versus temperature plots of as-prepared and annealed Tl-2223, Li$_{0.1}$, Li$_{0.2}$, and Li$_{0.4}$Tl-2223 samples. The $T_c$ (onset) is ~ 122 K for as-prepared Tl-2223, while the resistive transition for as-prepared Li$_{0.2}$Tl-2223 is ~ 125 K. The normal-state resistivity in Li$_{0.2}$ is high compared to Li$_{0.4}$Tl-2223 and decreases in both samples after annealing. There is no such change observed in Tl-2223 except that the transition becomes slightly broad after annealing. Figure 5(b) shows the temperature-dependent resistivity plots for Li$_{0.1}$Tl-2223 sample. The as-prepared sample shows zero resistivity at 86 K which increases to 113 K after annealing. The normal-state resistivity of the as prepared sample is high as compared to that of the annealed sample, however it follows the same temperature dependent slope as can be observed in the Figure inset. Such curvatures have also been observed in Hg-1223 sample where possible loss of Hg was held responsible for such behavior.[18] Broadening of the superconducting transition width for $Li_{0.1}$Tl-2223 may occur because of the weakly coupled grains. Along with behaving like a flux pinning site, $Li^+$ also has the ability to influence the hole carrier concentration and structural bonding, which may cause the change in $T_c$ and normal-state resistivity.[11]

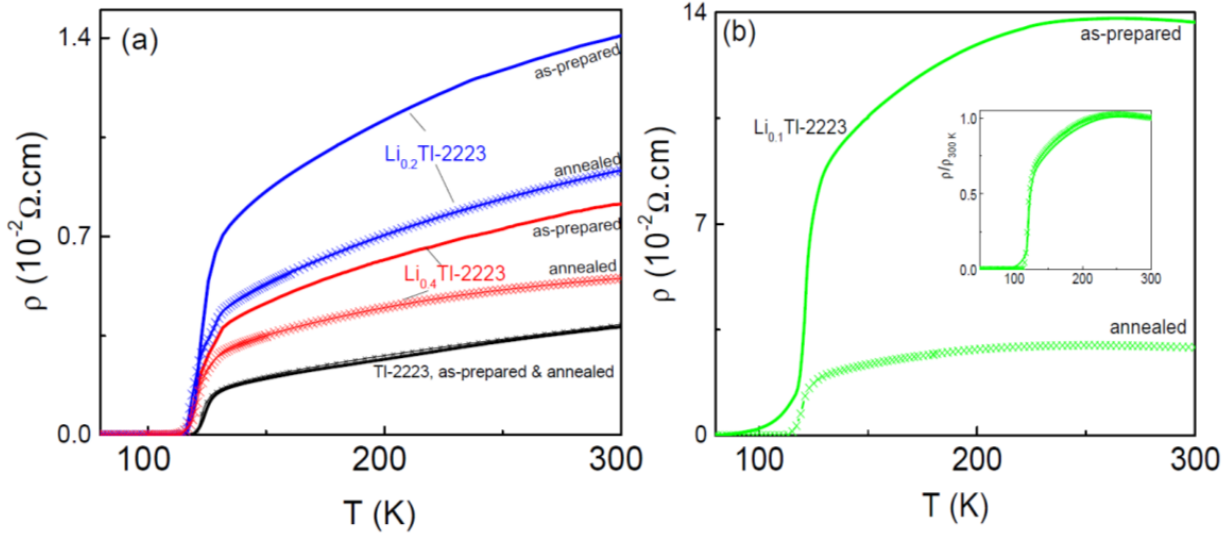

**Figure 5**: Temperature-dependent resistivity of as-prepared and annealed (a) Tl-2223, $Li_{0.2}$Tl-2223 and $Li_{0.4}$Tl-2223, and (b) $Li_{0.1}$Tl-2223. Inset in 5(b) is normalized data.

**Conclusions**

We have chemically substituted lithium in the structurally-complex $TlBa_2Ca_2Cu_3O_{9-\delta}$ cuprate superconductor, and shown that addition of $Li_2O$ and reduction of $Tl_2O_3$ reactants lower the synthesis temperature by 10 K. There is strong evidence, like the change in the lattice parameters and slight decrease in transition temperature, indicating that Li goes into the crystal structure of Tl-2223. Neutron diffraction technique may be employed for refinement of lighter oxygen and lithium atomic sites. It is interesting that nominal toxic thallium content in Tl-2223 can be chemically reduced by ~ 40% without much change in $T_c$, and higher levels may even be possible. In fact, lithium substitution can improve the superconducting volume fraction of known Tl-2223. We found that up to 10% $Li_2O$ can be added without any increase in the amount of impurity phases as compared to Li-free Tl-2223; although $Li_{0.2}$Tl-2223 showed the best $J_c$. The role of heavy thallium atom in high-temperature superconductivity may be realized through further substitution studies.

**Acknowledgments**

This research was primarily supported by the U. S. Department of Energy, Office of Science, Basic Energy Sciences, Materials Science and Engineering Division. RS was also supported by the National Science Foundation through grant No. DMR-0938330.